\documentclass[aps,twocolumn,floats,showpacs,superscriptaddress,floatfix]{revtex4}
\usepackage{epsfig}

\def\beginwide{
  \end{multicols} \vspace*{-0.5cm} \noindent
   \rule{3.5in}{.1mm}\rule{.1mm}{5mm} \widetext \medskip }
\def\beginwidetop{
  \end{multicols} \vspace*{-0.5cm} \noindent
        \widetext \medskip }
\def\endwide{
  \hspace*{3.35in}~\rule[-5mm]{.1mm}{5mm}\rule{3.5in}{.1mm}
  \begin{multicols}{2} \vspace*{-1.0cm} \noindent }
\def\endwidebottom{
  \begin{multicols}{2} \vspace*{-1.0cm} \noindent }

\begin{document}

\title{Ground state optimization and hysteretic demagnetization:\\
the random-field Ising model}

\author{Mikko J. Alava} 
\affiliation{Helsinki University of
Technology, Laboratory of Physics, HUT-02105 Finland} 

\author{Vittorio Basso} 
\affiliation{Istituto Elettrotecnico Nazionale Galileo Ferraris,  
strada delle Cacce 91, I-10135 Torino, Italy} 

\author{Francesca Colaiori}
\affiliation{INFM unit\`a di Roma 1 and SMC, Dipartimento di Fisica,
Universit\`a "La Sapienza", P.le A. Moro 2 00185 Roma, Italy}

\author{Lorenzo Dante} 
\affiliation{INFM unit\`a di Roma 1 and SMC, Dipartimento di
Fisica, Universit\`a "La Sapienza", P.le A. Moro 2 00185 Roma, Italy}
\affiliation{CNR Istituto di Acustica ``O. M. Corbino'', via del fosso
del Cavaliere 100, 00133 Roma, Italy} 

\author{Gianfranco Durin} 
\affiliation{Istituto Elettrotecnico Nazionale Galileo Ferraris, 
strada delle Cacce 91, I-10135 Torino, Italy} 

\author{Alessandro Magni}
\affiliation{Istituto Elettrotecnico Nazionale Galileo Ferraris, 
strada delle Cacce 91, I-10135 Torino, Italy} 

\author{Stefano Zapperi} 
\affiliation{INFM unit\`a di Roma 1 and SMC,
Dipartimento di Fisica, Universit\`a "La Sapienza", P.le A. Moro 2
00185 Roma, Italy}

\begin{abstract}
We compare the ground state of the random-field Ising model with
Gaussian distributed random fields, with its non-equilibrium
hysteretic counterpart, the demagnetized state.  This is a low energy
state obtained by a sequence of slow magnetic field oscillations with
decreasing amplitude. The main concern is how optimized the
demagnetized state is with respect to the best-possible ground state.
Exact results for the energy in $d=1$ show that in a paramagnet, with
finite spin-spin correlations, there is a significant difference in
the energies if the disorder is not so strong that the states are
trivially almost alike. We use numerical simulations to better
characterize the difference between the ground state and the demagnetized state.  
For $d\geq 3$ the random-field Ising model displays a disorder induced phase
transition between a paramagnetic and a ferromagnetic state.  The
locations of the critical points $R_c^{(DS)}$, $R_c^{(GS)}$ differ for
the demagnetized state and ground state.  Consequently, it is in this
regime that the optimization of the demagnetized stat is the worst
whereas both deep in the paramagnetic regime and in the ferromagnetic
one the states resemble each other to a great extent.  We argue based
on the numerics that in $d=3$ the scaling at the transition is the
same in the demagnetized and ground states. This claim is
corroborated by the exact solution of the model on the Bethe lattice,
where the $R_c$'s are also different.
\end{abstract}

\maketitle
\section{introduction}
         
The relation between equilibrium and non-equilibrium states is a
central problem in the physics of disordered systems. Disorder induces
a multitude of metastable states in which the system can easily be
trapped. The dynamics is usually very slow, or glassy, and on
observational timescales the system is basically always out of
equilibrium. On the other hand, from the theoretical point of view it
is easier to consider equilibrium properties, since in this case it is
possible to use all the machinery of statistical physics to tackle the
problem.  The question is whether the equilibrium properties of
disordered systems provide a faithful representation of the
non-equilibrium states in which the system is likely to be found in
practice. This dichotomy is at the core of many unsolved issues in the
field of disordered system. Typical quantities that one could compare
are the energy, the geometric characterization of the state (as
domains in magnets), and the energy cost of excitations.

A simplification of the problem is obtained considering only athermal
processes, in which the temperature of the system plays no role and
can be ignored.  The equilibrium state is in this case just the ground
state (GS), the state of minimal energy \cite{ALA-01}. A zero
temperature, non--equilibrium dynamics is purely relaxational: the
system falls simply in the closest metastable state. A convenient way
to allow the system to explore the various metastable states is by
applying an external magnetic field. Different field histories
typically result in hysteresis and lead to different metastable
configurations \cite{Bertotti}.

The demagnetization process consists in applying a slowly varying AC
field with decreasing amplitude, and provides a simple way to access
low energy states \cite{Bertotti}. It has been studied for more than a
century, but until recently the question how close the demagnetized
state (DS) is to the true GS was not addressed.  This is the concern
of our work, the problem of how such an optimization process works in
the case of a random magnet.  Recently, Pazmandy et al. have proposed
the demagnetization process as the basis for a new optimization
algorithm for disordered systems \cite{ZAR-02}.  The idea behind such
``hysteretic optimization'', is that demagnetization leads to a low
energy state, sufficiently close to the GS, which can then be reached
by applying other methods using the DS as an input. The method was
tested for different models like spin glasses and NP--hard problems.

Here we will concentrate on the random-field Ising model (RFIM), that,
while retaining some complex features characteristic of disordered
systems, still allows for a theoretical analysis \cite{NAT-00}. In the
RFIM, due to the absence of frustration, the equilibrium state is
relatively simple, however, the non--equilibrium dynamics is far from
trivial.  Due to the coupling of the local disorder to the order
parameter, even the GS presents a variety of phenomena, which can be
studied numerically \cite{OGI-86,HAR-99,HAR-01,MID-02}.  In fact the
GS of the RFIM can be found in a polynomial CPU-time, with exact
combinatorial algorithms \cite{ALA-01} and solved exactly in $d=1$ and
on the Bethe lattice \cite{BRU-84,SWI-94}. The equilibrium critical
exponents for random field magnets have been measured experimentally
in Fe$_{0.93}$Zn$_{0.07}$F$_2$ \cite{SLA-99,YE-02}.

The hysteretic properties of non equilibrium RFIM have been widely
studied in the recent literature.  The hysteresis loops display a
disorder induced phase transition: for low disorder the loop has a
macroscopic jump at the coercive field, while at high disorder the
loop is smooth, at least on the macroscopic scale
\cite{SET-93,DAH-96,PER-99}.  At smaller scale the magnetization curve
is highly discontinuous, showing Barkhausen-type bursts, in
correspondence to jumps between different metastable states
\cite{SET-01}. A disorder induced non-equilibrium phase transition in
the hysteresis loop has been studied experimentally in Co-CoO films
\cite{BER-00} and Cu-Al-Mn alloys \cite{MAR-03}.

Extensive numerical simulations have been used to characterize
disorder induced transitions in the non-equilibrium RFIM and critical
exponents have been estimated in several dimensions
\cite{PER-99,PER-03,PER-04}.  The model has been studied by the
renormalization group and the exponents have been computed in a
$\epsilon=6-d$ expansion \cite{DAH-96}.  In addition the hysteresis
loop has been computed exactly in $d=1$ and on the Bethe lattice,
where the disorder induced transition is present for sufficiently
large coordination number. While in $d=1$ there is definitely no
transition, the situation in $d=2$ is less clear.  Recently the
problem of minor loops has been tackled analytically and
numerically. In particular, the demagnetization curve has been
computed exactly in $d=1$ \cite{DAN-02} and on the Bethe lattice
\cite{COL-02}, extending previous calculations
\cite{SHU-96,SHU-00,DHA-97,SAB-00} of minor loops.

The equilibrium properties of the RFIM are governed by a
zero-temperature fixed point, and in finite dimensions ($d<5$ in
practice) GS calculations have elucidated the properties of the phase
diagram. In $d\geq 3$ the GS displays a ferromagnetic phase transition
induced by the disorder. As domain wall energy arguments and exact
mathematical results indicate, in $d=2$ there is no phase transition
but an effective ferromagnetic regime for small systems, while in
$d=1$ the RFIM is trivially paramagnetic. It has been suggested that
the transition in the GS is ruling the transition in the
non-equilibrium hysteresis loop, also because mean-field calculations
give the same results in and out of equilibrium
\cite{MAR-94}. Numerical values of the exponents are close but not
equal, but one must consider the difficulties in extrapolating values
from the finite size scaling \cite{MAR-94,SET-94}. More recently, the
question of the universality of the exponents, with respect to the
shape of the disorder distribution, was discussed in $d=3$
simulations, mean-field theory, and on the Bethe lattice
\cite{SOU-97,DUX-01,DOB-02}.

Below we report a detailed comparison of the zero temperature
equilibrium and non-equilibrium properties of the RFIM with Gaussian
distribution of the random fields.  We first analyze the problem in
$d=1$, where exact results can be obtained. The average value of the
energy is computed as a function of the disorder strength for the DS
and the GS. A direct comparison of the two values shows that for weak
disorder the differences become more substantial, while for strong
disorder, where each spin basically aligns with the random field, the
difference tends to vanish. Numerical studies using the same disorder
realizations reveal that the main difference between the two states
comes from the complete reversal of GS domains in the DS. This is also
visible in the overlap between the GS and DS.

We then study the $d=3$ case in which both paramagnetic and
ferromagnetic behavior exist.  The question of whether the transitions
appearing in the GS and in the hysteresis loop are universal has often
been debated in the literature \cite{MAR-94,SET-94}. At the mean-field
level it is not possible to distinguish the equilibrium and the
non-equilibrium case and the transition if thus trivially the same. In
addition, the $\epsilon$ expansion for the equilibrium and hysteretic
transitions is the same to all orders, but one should always consider
the possibility of non-perturbative corrections to the field
theory. Numerical simulations in $d=3$ indicate that the critical
exponents and the critical disorder in the two transitions are
reasonably close, but the numerical uncertainties do not allow for a
conclusive statement about their identity. Here we directly compare
the behavior of the GS and the DS in $d=3$ close to the disorder
induced phase transitions.  We show that while the non universal
critical parameter $R_c$ differs in the two cases, the universal
finite-size scaling curve for the order parameter can be collapsed on
the same curve. This suggests some kind of universality in the GS and
the DS transitions. The numerical simulations for the GS and DS are
done for the same disorder realizations for the both cases, for cubic
lattices of linear sizes $L=10,20,40,80$. The results are averaged
over several realizations of the quenched random fields.  In both
cases, we compute the average magnetization as a function of the
disorder width.

A difference in the location of the critical point for equilibrium and
non-equilibrium behavior of the same model may appear rather peculiar
and one could be tempted to ascribe it to finite size corrections. In
order to clarify this issue, we have solved exactly the model on the
Bethe lattice and compared the results for GS and DS.  While the
exponents, as expected, are the same, coinciding with the results of
mean-field theory, the critical disorder differs in the two cases.
Namely the transition in the DS occurs at a lower disorder value.
Thus there is an intermediate parameter region where the GS is
ferromagnetic but the DS is paramagnetic. In conclusion, the solution
on the Bethe lattice corroborates the picture obtained from
simulations in $d=3$. From the optimization viewpoint, the $d=3$ case
shows an intermediate phase of ``bad'' correspondence between the GS
and DS, exactly as in $d=1$. This however stops as the $R_c^{(DS)}$ is
approached: naturally if both the states are ferromagnetic the
optimization of the DS is much easier. To further explore the question
of universality of the two transitions in the GS and in the DS, we
have computed the distribution of the magnetization at the respective
critical point, $R_c^{(DS)}$ and $R_c^{(GS)}$ for different lattice
sizes.  The distributions can again all be collapsed into the same
curve.

Finally, we consider the question of when is it actually possible to
reach the exact GS via demagnetization. To this end, we consider a
reverse field history (RFH) algorithm that allows in principle to
construct a field history to get to the GS, if possible at
all. Studies of the $d=1$ case illuminate the difficulty of optimizing
since it turns out that for anything but very strong disorders $R$ the
probability to reach the GS rapidly decays to zero.

Our main conclusion is that, in general, demagnetization is not a good
technique for reaching states that are truly close to the equilibrium,
except in cases where the outcome is clearly similar from the very
beginning (FM states and PM states where the disorder is strong). This
holds for both the energy of the states and also for the spin
configurations. A simple formulation is that, since the DS is not
optimized well in terms of the locations of domain walls, it has an
excess random field (Zeeman) energy.

The paper is organized as follows: in section II we define the model
and discuss its numerical treatment. In sec. III we analyze the
one-dimensional case, analytically and numerically. Section IV is
devoted to the behavior around the disorder induced transition in
$d=3$ and on the Bethe lattice. Section V demonstrates the RFH
algorithm, together with numerical studies. Conclusions are reported
in section VI. An account of some of these results was briefly
reported in Ref.~\cite{COL-04}.

\section{The random-field Ising model}

In the RFIM, a spin $s_i = \pm 1$ is assigned to each site $i$ of a
 $d-$dimensional lattice. The spins are coupled to their
 nearest-neighbors spins by a ferromagnetic interaction of strength
 $J$ and to the external field $H$. In addition, to each site of the
 lattice it is associated a random field $h_i$ taken from a Gaussian
 probability $\rho(h)=\exp(-h^2/2R^2)/\sqrt{2\pi}R$, with variance
 $R$.  The Hamiltonian thus reads
\begin{equation} {\cal H} = -\sum_{\langle i,j \rangle}Js_i s_j -
\sum_i(H+h_i)s_i,\label{eq:rfim}
\end{equation}
where the first sum is restricted to nearest-neighbors pairs.

In this paper we will consider only the case of zero temperature, both
in equilibrium and out of equilibrium. The $T=0$ equilibrium problem
amounts to find the minimum of ${\cal H}$ for a given realization of
the random-fields (i.e. the GS) and then eventually perform the
thermodynamic limit. This problem has been solved exactly in a number
of simple cases, namely in $d=1$ and on the Bethe lattice, for
particular disorder distributions and studied numerically in generic
dimensions.

The RFIM GS is solvable in a polynomial CPU-time, with exact
combinatorial algorithms. For the one-dimensional case, the solution
can be found via a mapping to a ``shortest path problem''
\cite{1drfim} which effectively places the domain walls in optimal
positions, corresponding to the global minimum of ${\cal H}$. For
higher dimensions, one starts by noticing that finding the RFIM GS is
equivalent to the min-cut/max-flow problem of combinatorial
optimization.  This can be solved in a variety of ways.  We use a
so-called push-relabel variant of the preflow algorithm
\cite{eira}. Such methods, properly implemented, are in general
slightly sub-linear in their performance as a function of the number
of spins in the problem.

For the out of equilibrium case, we need to specify an appropriate
dynamics, ruling the evolution of the spins. We will consider the 
dynamics proposed in Ref.~\cite{BER-90} and used 
in Refs.~\cite{SET-93,DAH-96,PER-99} to study the hysteresis loop.
At each time step the spins align with the local field
\begin{equation}
s_i = \mbox{sign}(J\sum_j s_j  + h_i +H),
\label{EQ:2}
\end{equation}
until a metastable state is reached. This dynamics can be used 
to obtain the hysteresis loop. The system is started from a 
state with all the spin down $s_i=-1$ and then $H$ is ramped
slowly from  $H \to -\infty$  to $H \to \infty$. The limit of
$dH/dt \to 0$ can be conveniently obtained by increasing the
field precisely of the amount necessary to flip the first
unstable spin. A single spin flip increases the local field
of the nearest neighboring spins, generating an avalanche of flippings.
When the systems finds another metastable state, the field
is increased again. This dynamics obeys return-point memory \cite{SET-93}: 
if the field is increased adiabatically the magnetization only depends 
on the state in which the field was last reversed.  
This property has been exploited in $d=1$ \cite{SHU-00,DAN-02} 
and in the Bethe lattice \cite{SHU-01,COL-02}
to obtain exactly the saturation cycle and the minor loops.

The main hysteresis loop selects a series of metastable states, which
in principle are not particularly close to the ground state.  To
obtain low energy states, we perform a demagnetization procedure: the
external field is changed through a nested succession $H = H_0 \to H_1
\to H_2 \to ..... H_n...\to 0$, with $H_{2n}>H_{2n+2}>0$, $
H_{2n-1}<H_{2n+1}<0$ and $dH \equiv H_{2n}-H_{2n+2}\to 0$. A perfect
demagnetization can be performed numerically using the prescription
discussed above to obtain $dH/dt\to 0$. Such a perfect demagnetization
is quite expensive computationally and it is convenient to perform an
approximate demagnetization using $dH=10^{-3}$.  A comparison of the
states obtained under approximate and perfect demagnetization shows
negligible differences.

\section{Ground state and demagnetized state in one dimension}
\subsection{Exact results: ground state}
The GS energy can be computed exactly in $d=1$ using
transfer matrix methods \cite{BRU-84}
The free energy of a chain of length $N$ is given by
\begin{equation}
F_N=-\frac{1}{\beta}\ln(Z_N)=-\frac{1}{\beta}\ln(Z_N^{+}-Z_N^{-})
\simeq-\frac{1}{\beta}\ln(Z_N^{+}Z_N^{-})
\label{F_N}
\end{equation}
where $Z_N$ is the partition function with free boundary conditions and 
$Z_N^{\pm}$ are the partition functions with the spin at site $1$ fixed up(down). 
These functions satisfy the following recursive relation:
\begin{equation}
Z_N^{\pm}=e^{\pm\beta h_1}(Z_{N-1}^{+}e^{\pm\beta J}+Z_{N-1}^{-}e^{\mp\beta J})
\label{Z_N}
\end{equation}
The last step in eq.(\ref{F_N}) uses the approximation 
$Z_N^+ + Z_N^- \simeq \sqrt{Z_N^+Z_N^-}$ which holds in the large 
$N$ limit since $Z_N^{\pm}$ both diverge with the ratio $Z_N^+/Z_N^-$ 
being finite.  From Eq. (\ref{Z_N}) it follows
\begin{equation} 
Z_N^{+}Z_N^{-}=Z_{N-1}^{+}Z_{N-1}^{-}(2 \cosh(\beta J)+2 \cosh(2 \beta x_N))
\label{ZZ}
\end{equation}
where $x_N=\frac{1}{2\beta}\ln(Z_N^+/Z_N^-)$, which gives for the total free energy
\begin{equation} 
F_N=F_{N-1}-\frac{1}{2\beta}\ln(2\cosh(\beta J)+2 \cosh(2 \beta x_N))
\label{F}
\end{equation}
where 
$x_N=\frac{1}{2 \beta} \ln (Z_N^+/Z_N^-),$
so that one can define a free energy per site
\begin{equation} 
f=-\frac{1}{2\beta}\ln(2\cosh\beta J+2 \cosh(2 \beta x_N)) .
\label{f}
\end{equation}
$x_N$ is a stochastic quantity satisfying the equation
\begin{equation} 
x_N=h_1+g(x_{N-1})
\label{x2}
\end{equation}
where $g(x)=\frac{1}{2\beta}\ln \left(\left(e^{2\beta(x+J)}+1\right)/
\left(e^{2\beta(x)}+e^{2\beta(J)}\right)\right)$
When $R \rightarrow 0$ Eq. (\ref{x2}) has a fixed point solution of 
$x_\infty=g(x_\infty)$. 
It is easy to check that $x_\infty=0$ is the only solution for any $J$ and 
$\beta$ finite, 
corresponding to the absence of a phase transition. 

When $R$ is non-zero $x_N$ is a random variable with an associated distribution 
$W_N(x)$, where 
\begin{equation}
W_N(x) dx=\mbox{Prob}(x<x_N<x+dx).
\label{W}
\end{equation}
$W_N(x)$ satisfies the recursive functional equation
\begin{eqnarray}
&W_{N+1}(x)=\int_{-\infty}^{\infty} dh P(h) \times &
\nonumber
\\
&\int_{-\infty}^{\infty} dx_1 W_N(x_1)\delta(x-h-H-g(x_1)) &
\label{W_N}
\end{eqnarray}
so that in the thermodynamic limit $W_{\infty}$ is given by the fixed point equation 
\begin{equation} 
W_{\infty}(x)=\int_{-\infty}^{\infty} dx_1 W_{\infty}(x_1) P(x-h-H-g(x_1))
\label{W_inf} \,.
\end{equation}
Once $W_{\infty}$ is known, any thermodynamic quantity can be computed. In particular, 
the free energy per spin, which is given by
\begin{equation} 
\langle f \rangle=-\frac{1}{\beta}\int_{-\infty}^{\infty}\! dx 
W_{\infty}(x)\left(\cosh(2\beta)+\cosh2\beta x\right) . 
\label{<f>}
\end{equation}
The magnetization at a site $0$ of an infinite lattice, is given by 
 \begin{eqnarray}
&\langle s_0 \rangle=\frac{Z^{\uparrow}-Z^{\downarrow}}{Z^{\uparrow}+Z^{\downarrow}}= &
\nonumber
\\
&\frac{\sqrt{Z^{\uparrow}/Z^{\downarrow}}-
\sqrt{Z^{\downarrow}/Z^{\uparrow}} }
{\sqrt{Z^{\uparrow}/Z^{\downarrow}}+\sqrt{Z^{\downarrow}/Z^{\uparrow}} }=
\tanh\left(\frac{1}{2}\ln(Z^{\uparrow}/Z^{\downarrow})\right) \,,&
\label{s0}
\end{eqnarray}
where $Z^{\uparrow\downarrow}$ are respectively the partition functions
with the spin at $0$ fixed up (down). 
These are given by 
\begin{equation} 
Z^{\uparrow\downarrow}\!=e^{\pm \beta h_0}
(e^{\pm \beta J}Z_r^+ +e^{\mp \beta J}Z_r^-)
(e^{\pm \beta J}Z_l^+ +e^{\mp \beta J}Z_l^-)
\label{Zupdown}
\end{equation}
where $Z^{\pm}_{r,l}$ are the partition functions for the semi-infinite right(left) 
lattice, with the spin at site $1$ ($-1$) fixed up(down). This gives 
\begin{equation} 
\langle s_0 \rangle=
\tanh(\beta(h_0+g(x_r)+g(x_l))).
\label{s02}
\end{equation}
Finally, The magnetization for the infinite lattice is obtained averaging 
over the quenched variables $x_{r,l}$:
\begin{eqnarray}
&m=\int_{-\infty}^{\infty} dh P(h)
\int_{-\infty}^{\infty} dx_r W_N(x_r)&
\nonumber
\\
&\int_{-\infty}^{\infty} dx_l W_N(x_l)
\tanh\left(\beta(h+g(x_r)+g(x_l))\right).&
\label{mag}
\end{eqnarray}

\subsection{Exact results: Demagnetized state}

In $d=1$ the magnetization and the energy per spin as a function of
the external field can be derived explicitly through a probabilistic
reasoning. We show how to get these results on the saturation loop,
focusing on the lower branch.  (The results on the upper branch can be
obtained by symmetry considerations.)  Similar but much more involved
reasoning can be repeated for any minor loop.

The central quantity to consider, in order to solve for the magnetization
as a function of the external field $H$ on the hysteresis loop, is the 
conditional probability for a spin to be up, conditioned to one of its 
nearest neighbors being down. To calculate this quantity, one can reason 
as follows:
fix the spin at site $i-1$ down. Define $p_m(H)$ as the probability for a 
spin to be up, given that exactly $m$ ($ m=0,1,2$) of its neighbors are up:
\begin{equation}
p_m(H)=P(h_{i}^{\mbox{{\tiny eff}}}>0)=\int_{\mbox{{\tiny
$(z-2m)J-H$}}}^{\infty}dh \rho(h) \,,
\label{pm}
\end{equation}
where $z$ is the coordination number ($z=2$ in $d=1$). 
Fix for a moment the spin at site $i$ down as well and look at the spin at site $i\!+\!1$. 
It will be up with probability $U_0$ and down with probability $1-U_0$. The spin at 
site $i$ will flip up with probability $p_1$ when the spin at $i\!+\!1$ is up and $p_0$
when it is down. Ultimately, the spin at $i$ will be up (conditioned to the spin at 
$i\!-\!1$ being down) with probability $U_0=U_0 p_1+(1-U_0)p_0$. It follows
\begin{equation}
U_0=\frac{p_0}{1-p_1+p_0}.
\label{p0}
\end{equation}
Once $U_0$ is known, a similar reasoning leads to the (unconditioned) 
probability $p(H)$ for a spin to be up: Fix the spin at site $i$ down. 
The spin at site $i\! -\! 1$ will be up with probability $U_0$ and down with 
probability $1-U_0$. The same holds for the spin at site $i \!+\!1$. 
Thus
\begin{equation}
p(H)=U_0^2 p_2+2U_0(1-U_0)p_1+(1-U_0)^2 p_0 ,
\label{p(h)}
\end{equation}
from which the magnetization is obtained as  $m(H)=2 p(H)-1$.

\begin{figure}[ht]
\centerline{\psfig{file=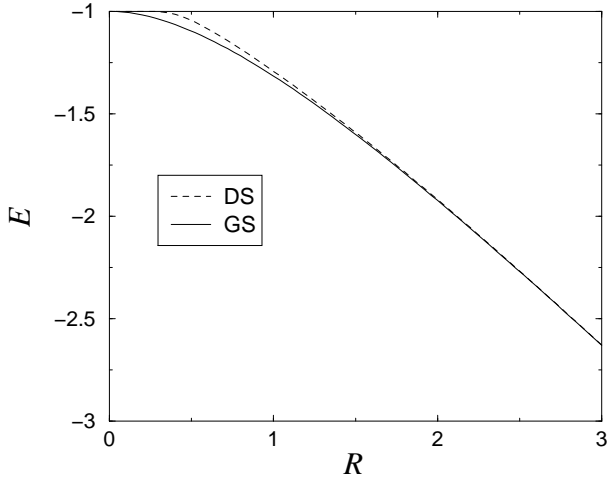,width=8cm,clip=!}}
\caption{The energy of the GS is compared with the one of 
DS. The values are computed exactly in $d=1$ as a function of
the disorder width $R$.}
\label{fige1d}
\end{figure}

\begin{figure}[ht]
\centerline{\psfig{file=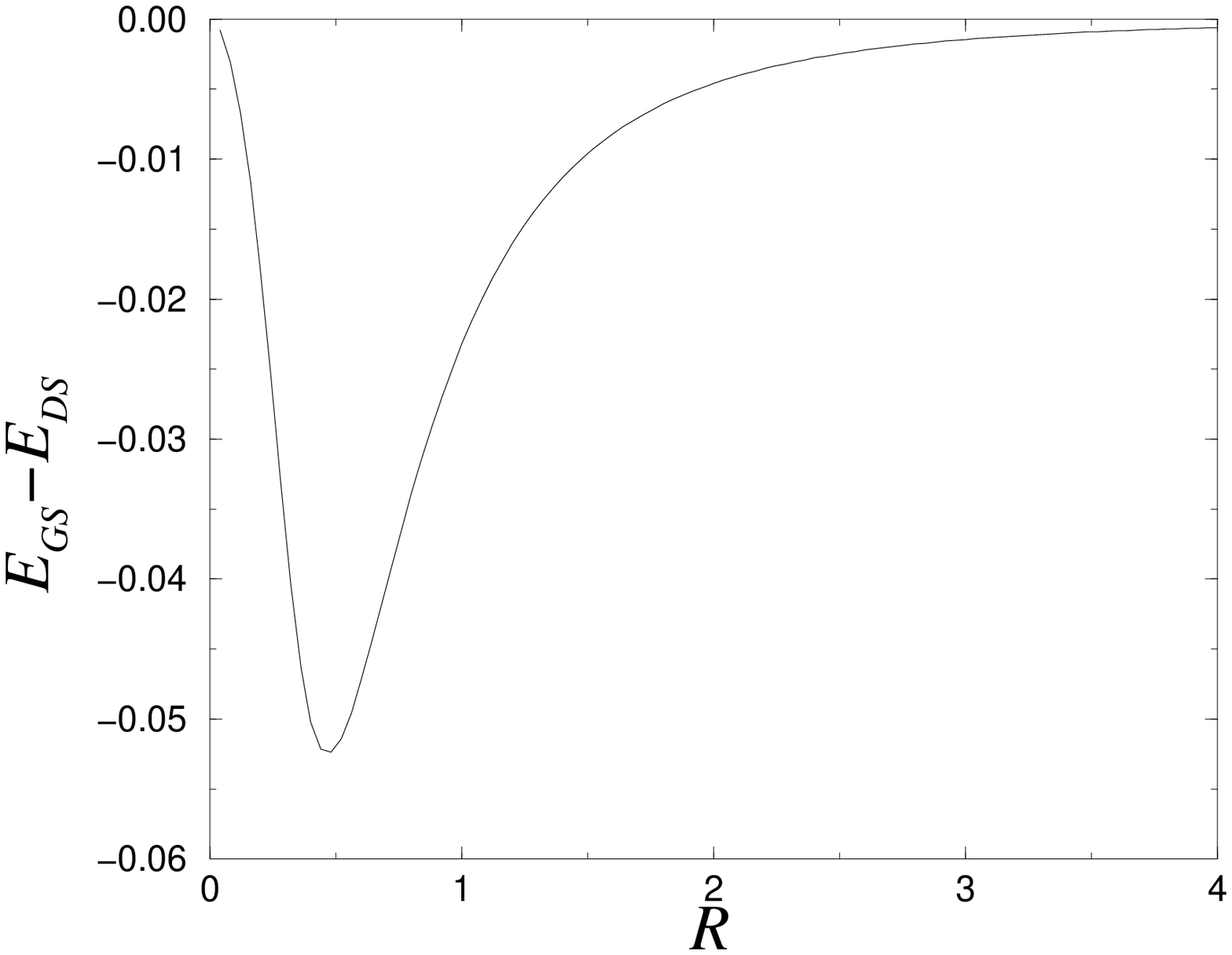,width=8cm,clip=!}}
\caption{The energy difference between  
the GS and the DS computed exactly in $d=1$ as a function of
the disorder width $R$.}
\label{figde1d}
\end{figure}

The energy per spin on the saturation loop is obtained as follows. Due to translational 
invariance: 
\begin{equation}
E=\frac{\langle {\cal H}\rangle}{N}=-J\langle s_i s_{i+1}\rangle
-H\langle s_i\rangle -\langle h_{i}s_i\rangle.
\label{energy}
\end{equation}
To calculate the spin--spin correlation $\langle s_i s_{i+1}\rangle$ 
we introduce the probabilities $\Phi^{\mbox{\tiny $++$}}$, 
$\Phi^{\mbox{\tiny $+-$}}$, $\Phi^{\mbox{\tiny $-+$}}$, 
$\Phi^{\mbox{\tiny $--$}}$ for adjacent spins to be respectively 
up--up, up--down, down--up, and down--down. These quantities are not independent, 
since they have to satisfy the obvious identities: 
$\Phi^{\mbox{\tiny $+-$}}\!=\Phi^{\mbox{\tiny $-+$}}$, 
$\Phi^{\mbox{\tiny $++$}}\!+\Phi^{\mbox{\tiny $+-$}}\!=p(H)$, and 
$\Phi^{\mbox{\tiny $--$}}\!+\Phi^{\mbox{\tiny $+-$}}\!=\!1\!-\!p(H)$. 
Thus it is sufficient to calculate one of them, for example $\Phi^{\mbox{\tiny $--$}}$. 
This is done by separating the four contributions from the possible boundary 
conditions determined by the values of the spins at sites $i\!-\!1$ and $i\!+\!2$: 
When they are both down, the probability for the couple of spins at sites 
$i$ and $i\!+\!1$ to be both down is $U_0^2(1-p_1(H))^2$, 
when one is up and the other one is down it is $2U_0(1-U_0)(1-p_1(H))(1-p_0(H))$, 
and when both of them are up it is $(1-U_0)^2(1-p_0(H))^2$.   
Adding up the four contributions one gets $\Phi^{\mbox{\tiny $--$}}=(1-U_0)^2$.   
This fixes the other probabilities to be 
$\Phi^{\mbox{\tiny $+-$}}\!=\Phi^{\mbox{\tiny $-+$}}=2p\!-\!1\!+\!(1-U_0)^2$, and 
$\Phi^{\mbox{\tiny $++$}}=1\!-\!p-(1-U_0)^2$. Thus, the spin--spin correlation is 
\begin{equation}
\langle s_i s_{i+1}\rangle=
\Phi^{\mbox{\tiny $++$}}\!\!+\Phi^{\mbox{\tiny $--$}}
\!-2\Phi^{\mbox{\tiny $+-$}}\!\!=
4\left(p-(1-U_0)^2\right)-3 \,.
\label{correlation}
\end{equation}
The average value $\langle h_i s_i\rangle$
can be obtained by averaging over the field $h^{'}$ the product of $h^{'}$ times the 
average value of the spin $s_i$ over the local fields other then $h_i$, 
once the field at $i$ is fixed at the value $h^{'}$:  
\begin{equation}
\langle h_i s_i\rangle=\int_{-\infty}^{+\infty}dh^{'}\rho(h^{'})h^{'}
\langle s_i | h^{'} \rangle .
\label{ehi} 
\end{equation} 
The conditional average $\langle s_i|h^{'}\rangle$ is given by 
$2p(H|h^{'})\!-\!1\!$ where $p(H|h^{'})$ is the conditional probability for a spin to be up at 
an external field $H$, given that its local random field is fixed at the value 
$h^{'}$. From Eq. (\ref{p(h)}) this is trivially given by 
\begin{eqnarray}
p(H|h^{'}) &=& U_0^2 \theta(h^{'}+H+2J)\nonumber\\
           &+& 2U_0(1-U_0) \theta(h^{'}+H)\nonumber\\
           &+& (1-U_0)^2 \theta(h^{'}+H-2J) \,,
\end{eqnarray} 
which finally gives
\begin{eqnarray}
\langle h_i s_i\rangle &=& 2 U_0^2 \int_{\mbox{\tiny $-H-2J$}}^{+\infty}dh^{'} \rho(h^{'}) h^{'}\nonumber\\
                       &+& 4U_0(1-U_0) \int_{\mbox{\tiny $-H$}}^{+\infty}dh^{'} \rho(h^{'}) h^{'}\nonumber\\
                       &+& 2(1-U_0)^2 \int_{\mbox{\tiny $-H+2J$}}^{+\infty}dh^{'} \rho(h^{'}) h^{'}-\bar{h^{'}} .
\end{eqnarray}
In particular, for a Gaussian distribution with $\bar{h^{'}}=0$ and variance 
$R$ the integrals can be performed analytically and the result is 
\begin{eqnarray}
\langle h_i s_i\rangle&=&
\sqrt{\frac{2}{\pi}}Re^{-\frac{H^2}{2R^2}}\left[
2U_0^2e^{\frac{2J}{R^2}}\cosh\left(2JH/R^2\right) \right.
\nonumber\\
&+&\left. e^{2J(J-\frac{H}{2R^2})}(1-2U_0^2)+2U_0(1-U_0)  \right] .
\end{eqnarray}
The energy per site on the lower branch of the saturation loop is in general 
given by 
\begin{eqnarray}
E(H)\!\!\!\!&=\!\!&-4J\left(p(H)-(1-U_0)^2\right)+3J-H(2p(H)-1)
\nonumber\\
&-&
2 U_0^2 \int_{\mbox{\tiny $-H\!-\!2J$}}^{+\infty}dh^{'} \rho(h^{'}) h^{'}\nonumber\\
&+&
4U_0(1-U_0)  \int_{\mbox{\tiny $-H$}}^{+\infty}dh^{'} \rho(h^{'}) h^{'}
\nonumber\\
&+& 2(1-U_0)^2 \int_{\mbox{\tiny $-\!H\!+\!2J$}}^{+\infty}dh^{'} \rho(h^{'}) h^{'}-\bar{h^{'}} .
\end{eqnarray}

Similar but much more involved reasonings can be repeated for any minor loop --
eventually for a series of nested loops leading to the demagnetized state --
providing a series of recursive equations for the magnetization, the 
spin--spin, and the spin--field correlations, which are the quantities needed 
to compute the energy. 
If the external field is changed through a nested succession
$H =H_0 \to H_1 \to H_2 \to ..... H_n...\to 0$, with  
$H_{2n}>H_{2n+2}>0$, $ H_{2n-1}<H_{2n+1}<0$ and $dH\equiv H_{2n}-H_{2n+2}\to 0$,  
the spin--spin correlations are given recursively by 
\begin{eqnarray}
&\langle s_i s_{i\!+\!1}\rangle_{H_{2n}}\!-\langle s_i
s_{i\!+\!1}\rangle_{H_{2n\!-\!1}}\!\!=\!\!&\nonumber\\ &
4U^2_{2n}\left(p_2(H_{2n})-p_2(H_{2n\!-\!1})\right) & \nonumber\\ &
-4D^2_{2n\!-\!1}\left(p_0(H_{2n})-p_0(H_{2n-1})\right)&
\end{eqnarray}
where $U_k$ and $D_k$ are respectively the probabilities for a spin to be 
up(down) conditioned to one of its neighbors being down, and satisfy in turn 
a set of recursive equations. Similar equations hold for magnetization and 
spin--field correlation, leading to a complicated recursive formula for the energy.   
The results of such calculations are shown in 
Figs. (\ref{fige1d},~\ref{figde1d}), 
where the energy of the demagnetized state is compared with the energy of the 
ground state evaluated in the previous section.  
 
\subsection{Simulations: how optimized is the demagnetized state?}

In one dimension the comparison of the DS and the GS is the easiest
since the domain walls are just point-like.  For the GS we know that
it is optimized such that all the large enough local random field
fluctuations nucleate domains of the same sign. The rest of the random
landscape is split up into regions that align themselves with such
fluctuations depending on the sign of the random field excess,
$\sum_{i \in \mathrm{region}} h_i$. As a result the Zeeman energy of
domains is linear in domain size, $E_Z \sim l_d$, and the asymptotic
mean domain length follows the Imry-Ma prediction $\langle l_{GS}
\rangle \sim 1/R^2$. Moreover since the random landscape has a finite
correlation length the domain size distribution is exponential
\cite{1drfim}.

\begin{figure}[htb]
\centerline{
\psfig{figure=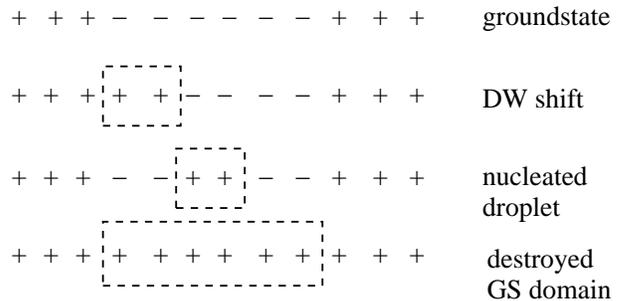,width=8cm} }
\caption{An illustration of the possible 
mechanisms for the deviations between GS and DS.}
\label{mechanisms}
\end{figure}

Any qualitative differences in the DS will follow from three separate
mechanisms: 1) shifts of domain walls, 2) creation of domains inside
intact GS domains and 3) destruction of GS domains
(Fig.~\ref{mechanisms}). From the point of view of ''optimization" the
first one is of trivial concern, since it would have little effect
e.g. on the scaling of $E_{Z,DS}$. The second one is more detrimental
if the energy difference to the GS is considered. In addition to the
cost of the two domain walls it subtracts a contribution from the
Zeeman energy of the domain that persists and surrounds (in the GS)
the one that is not created in the DS. The third one would make the
largest change to the total energy, since for $l_d \gg 1$ the energy
of a domain consists mostly of its Zeeman energy.

\begin{figure}[htb]
\centerline{
\psfig{angle=270,width=8cm,figure=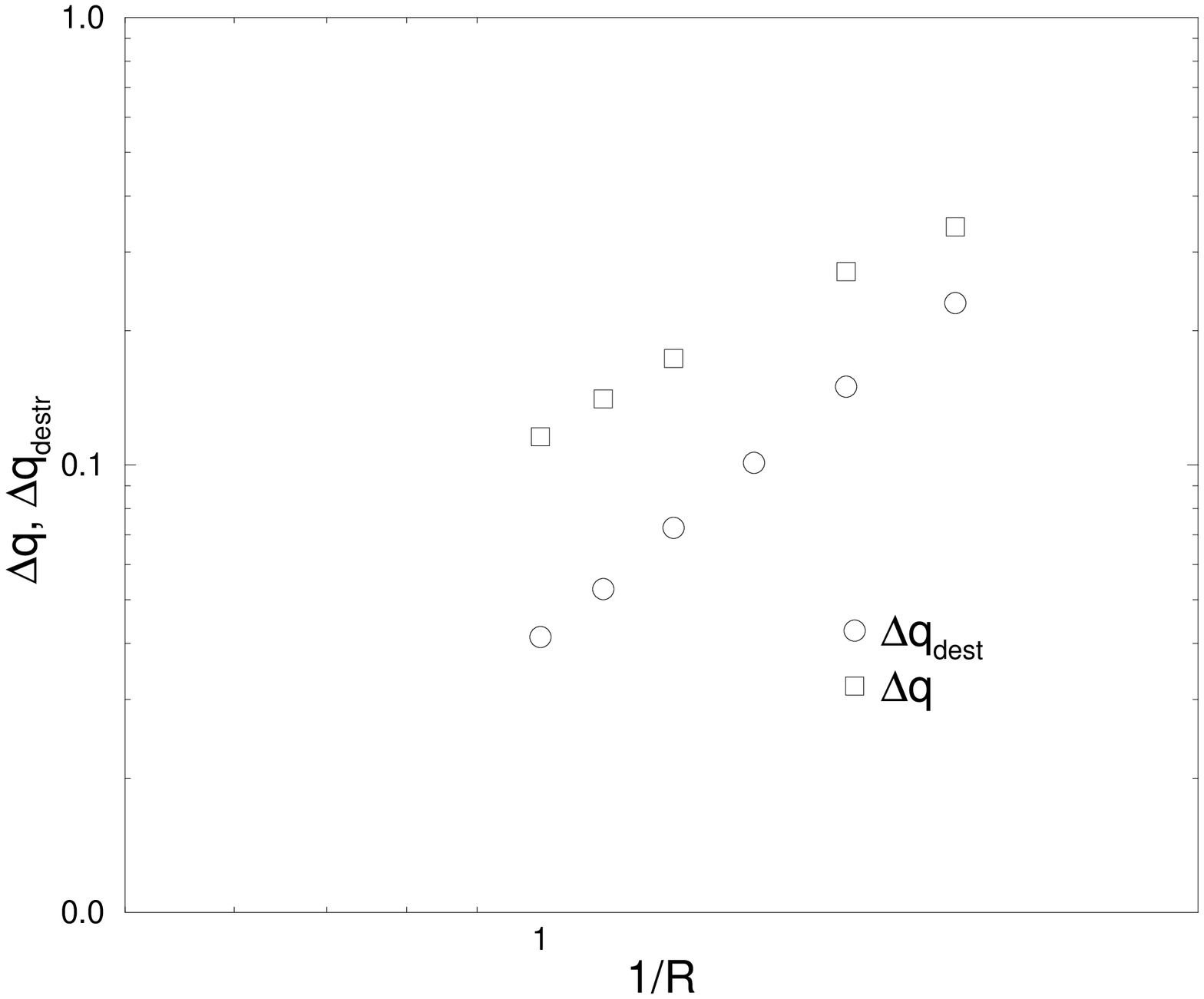} }
\caption{The average change in the spin-spin
overlap between the GS and the DS
($\Delta q$) and the contribution to that 
from completely ``destroyed'' GS domains
($\Delta q_{destr}$), as a function of $R$.
  }
\label{qs}
\end{figure}

\begin{figure}[htb]
\centerline{
\psfig{angle=270,width=8cm,figure=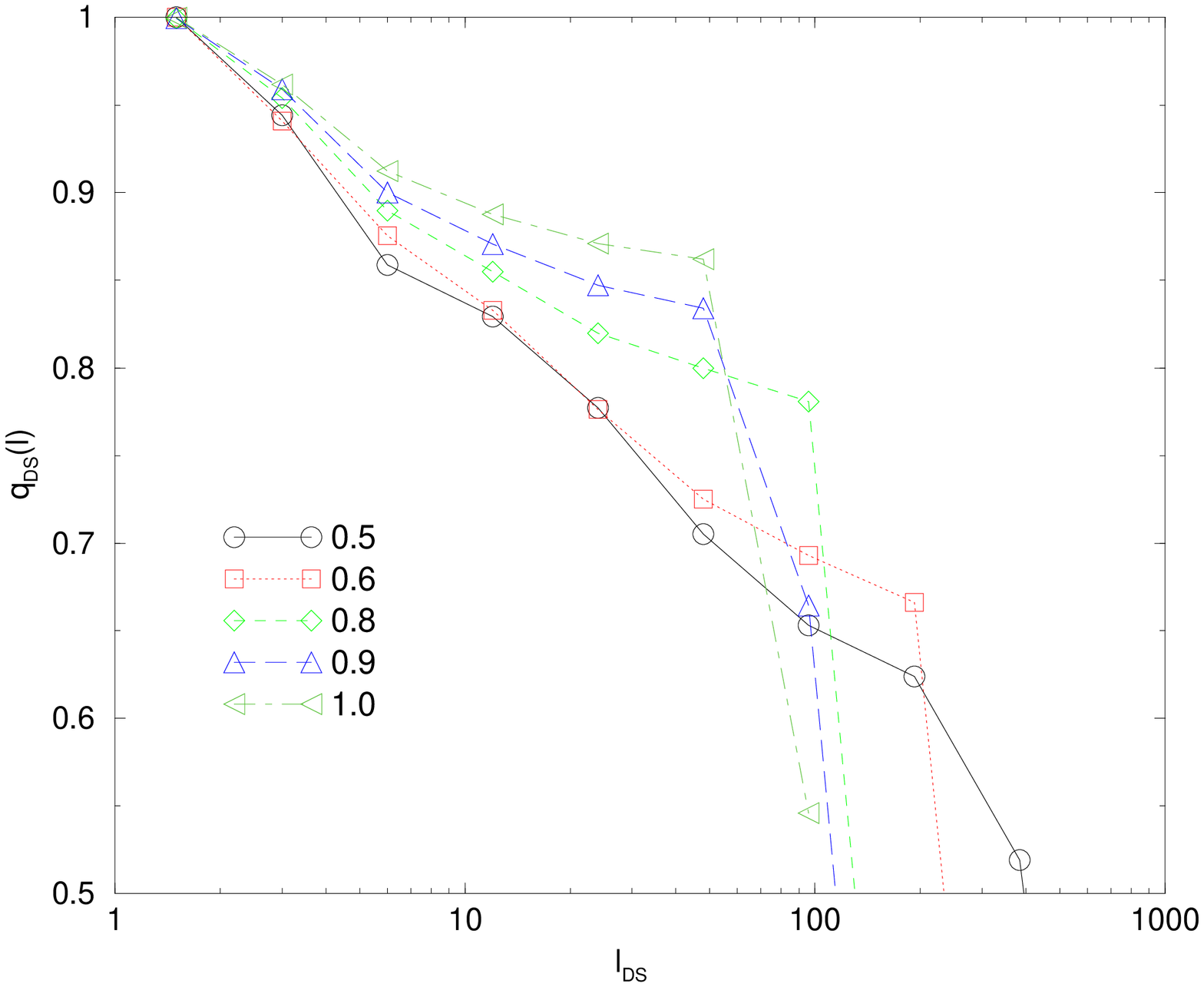} }
\caption{The average overlap of a DS domain of size $l$
with the GS domain spin state at the same locations for
$R=$ 0.5, 0.6, 0.8, 0.9, 1.0. The overlap decreases with
$l_{domain,DS}$.
  }
\label{dsoverlap}
\end{figure}

\begin{figure}[htb]
\centerline{
\psfig{angle=270,width=8cm,figure=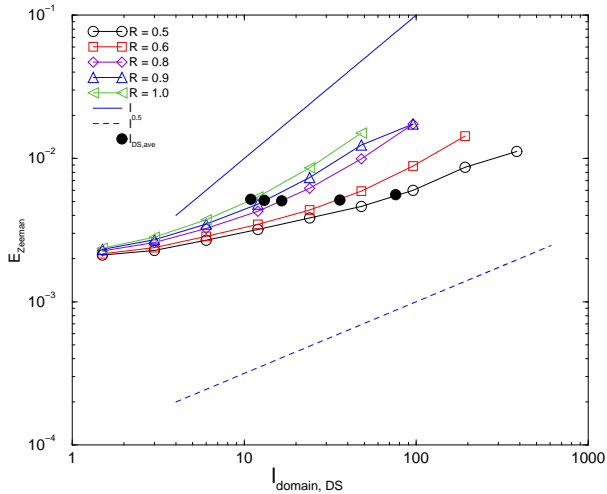} }
\caption{The Zeeman energy of DS domains as a function
of the size, $l_{domain,DS}$. The black circles mark the
average DS domain size for a given $R$. The two lines
above and below the data
indicate optimal, linear (GS-like) scaling and the Imry-Ma-like
$l^{1/2}$-scaling, respectively.
  }
\label{dszeeman}
\end{figure}

Numerical studies of the DS domain structure indicate that with
decreasing $R$ the average domain size increases faster than in the
GS, while the size distribution $P(l_d)$ remains exponential. This is
accompanied by a reduction in the overlap $q=(\langle \sigma_{GS}
\sigma_{DS} \rangle +1)/2$ between these two states. For $R$ large the
overlap is close to unity; strong local fields $h_i$ align the spins
in the same way regardless of the mechanism by which the spin state is
created. For $R$ small the local field is no longer strongly
correlated with the orientation of the spin, and thus whether the GS
and DS are locally aligned depends on how optimized the latter is.

The fundamental mechanism for the deviations between the states seems
for $R$ small to be the ``destruction" of GS domains (see
Fig.~\ref{mechanisms} again). This is demonstrated in Fig.~\ref{qs} by
depicting the change $\Delta q$ in the overlap that comes solely from
missing GS domains.  The conclusion from this dominance is that the
demagnetized states typically miss regions in which the integrated
field fluctuation is large which as such leads in the GS to the
formation of GS domain. Therefore the overlap should get smaller the
larger the scale-length on which one compares the DS and GS is, and
this is confirmed by Fig.~\ref{dsoverlap} which shows the overlap
between a DS domain and the GS as a function of the length of the DS
domain.

The importance of such destroyed domains can also be seen in the total
contribution to the energy difference between the DS and GS. For $R$
small this is again dominated by the missing GS domains. In general
the difference between the energies of the GS and DS derives from the
combination of domain walls and Zeeman energy. Fig. \ref{dszeeman}
shows that for $l_d$ small the DS domains do not have much Zeeman
energy.  This changes if $l_d$ is larger, in which regime the scaling
approaches the Imry-Ma -like scaling ($l_d^{0.5}$). The implication is
that the field energy of large domains in the DS self-averages, and
comes from a sum of random contributions (ie. the domains contain
regions where the actual random field sum is {\em opposite} to the
spin orientation, such as the missing GS domains).  The cross-over
between the small $l_d$-behavior and the asymptotic scaling is located
close to $\langle l_d \rangle_{DS}$.

\section{Around the disorder induced transition} 
\subsection{Simulations in $d=3$}
 
The RFIM displays a disorder induced phase transition both in the GS
and in the hysteresis loop, which can also be observed analyzing the
DS \cite{DAN-02,COL-02,CAR-03}.  If the GS and the DS are always
paramagnetic, the transition is absent and thus we perform numerical
simulations in $d=3$.  Our aim is to characterize the difference
between DS and GS around the disorder induced transition.

In $d=3$ for low disorder, the GS is ferromagnetic, while for higher
disorder it becomes paramagnetic. For Gaussian disorder, the
transition point has been located numerically at $R_c^{(GS)}\simeq
2.28$.  It is possible to define the usual set of critical exponents
characterizing the phase transition and compute the values by exact GS
calculations. For instance, the magnetization $M\equiv \langle |m|
\rangle$, with $m\equiv \sum_i s_i/N$, scales close to the transition
point as
\begin{equation}
M=A r^\beta,
\end{equation}
where $r\equiv (R-R_c)/R_c$ is the reduced order parameter and $A$ is
a non-universal constant.  The correlation length defines another
exponent $\xi=(B r)^{-\nu}$ --where $B$ is another non-universal
constant-- which rules the finite size scaling of the model
\begin{equation}
M=A L^{-\beta/\nu}f\left(B L^{1/\nu}(R-R_c)/R_c\right).
\label{eqfss}
\end{equation}
Simulations yield $\nu^{(GS)} \simeq 1.17$ and $\beta^{(GS)}=0.02$.

A disorder induced transition is also found in the hysteresis loop. At
low disorder the loop shows a macroscopic jump, which disappears at a
critical value for the disorder. This transition reflects itself in
the DS, which is ferromagnetic when the main loop has a jump and is
paramagnetic otherwise. The transition point has been obtained
numerically in $d=3$ as $R_c^{(DS)} \simeq 2.16$ and the critical
exponents have been measured. In particular, Ref. \cite{CAR-03}
reports data collapses with $\beta_{(DS)}=0.04$ and
$\nu_{(DS)}=1.41$. While there is strong evidence that the exponents
measured in the DS should be equal to those measured on the main loop,
the relation with the equilibrium transition is not clear.

We notice first that numerical simulations reported in the literature
indicate that the transition appears at slightly different locations
in the GS and in the DS. Hartmann and Nowak report $R_c^{GS}=2.29\pm
0.04$ for the GS with system size up to $L=80$, Hartmann and Young
refine this value to $R_c^{(GS)}=2.28\pm 0.01$ with sizes up to
$L=96$, which is also confirmed by Middleton and Fisher which estimate
$R_c^{(GS)}=2.27\pm0.04$. For the hysteresis loop the best estimate is
$R_c=2.16 \pm 0.03$, with system sizes up to $L=320$ and a similar
value for the $DS$ \cite{DAN-02,CAR-03}.  Thus, unless strong finite
size effects take place, one is tempted to conclude that the two
transitions take place at two different values of $R_c$. 

\begin{figure}[ht]
\centerline{\psfig{file=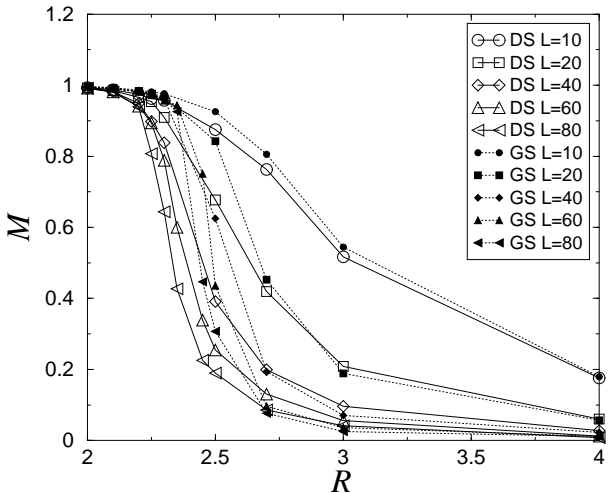,width=8cm,clip=!}}
\caption{The magnetization of the GS and the DS in $d=3$ for different
system sizes $L$ and disorder $R$.}
\label{figmgsds}
\end{figure}

\begin{figure}[ht]
\centerline{\psfig{file=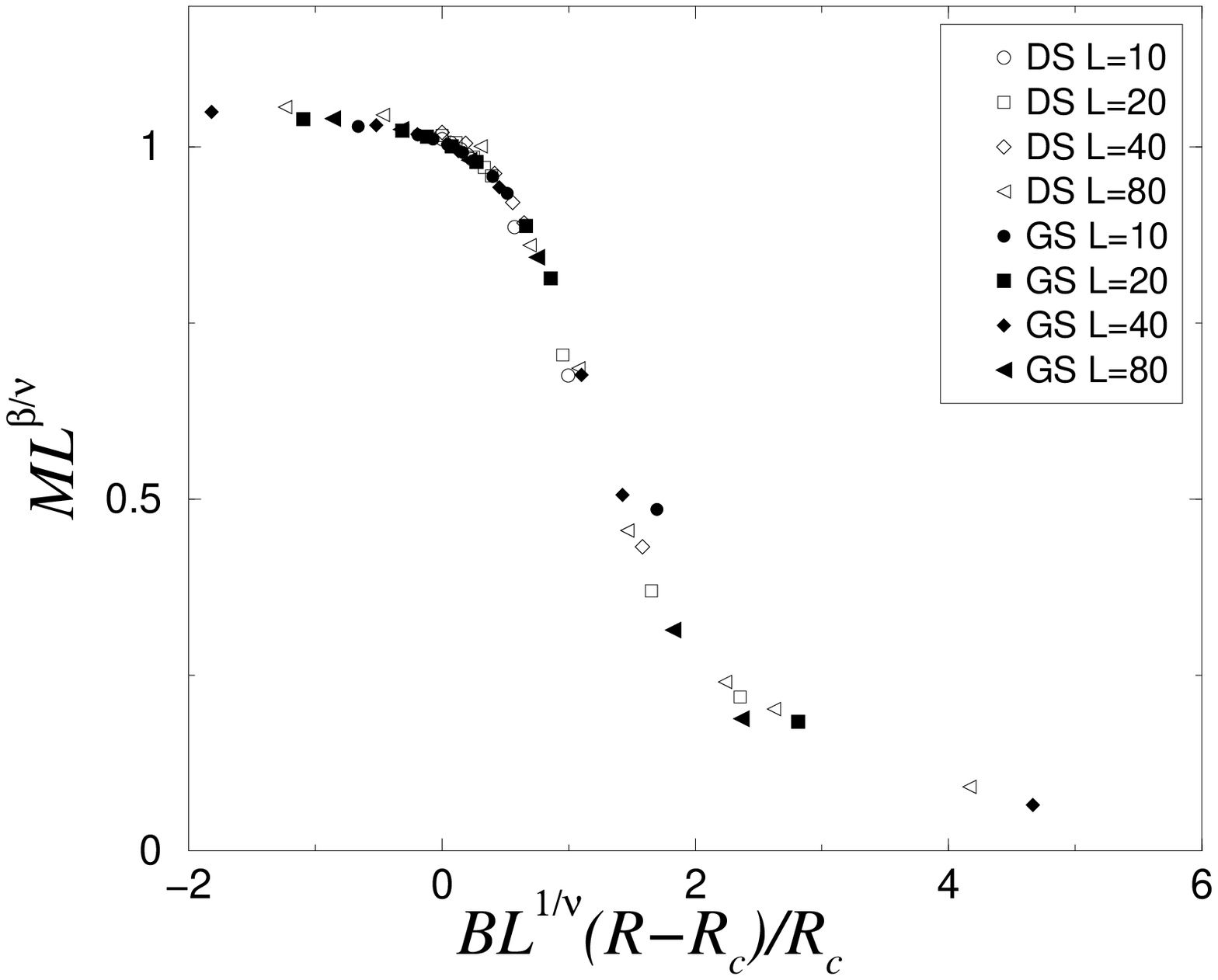,width=8cm,clip=!}}
\caption{Numerical results in $d=3$: 
The magnetization can be collapsed using  
$R_c=2.28$ (GS) and $R_c=2.16$ (DS), $1/\nu=0.73$ and $\beta=0.03$. 
The scaling curve is the same for DS and GS indicating universal behavior. 
The values for the ratios of the non-universal constants are 
$A_{DS}/A_{GS}= 1$ and $B_{DS}/B_{GS} = 0.68$.}
\label{figcollapse}
\end{figure}

\begin{figure}[ht]
\centerline{\psfig{file=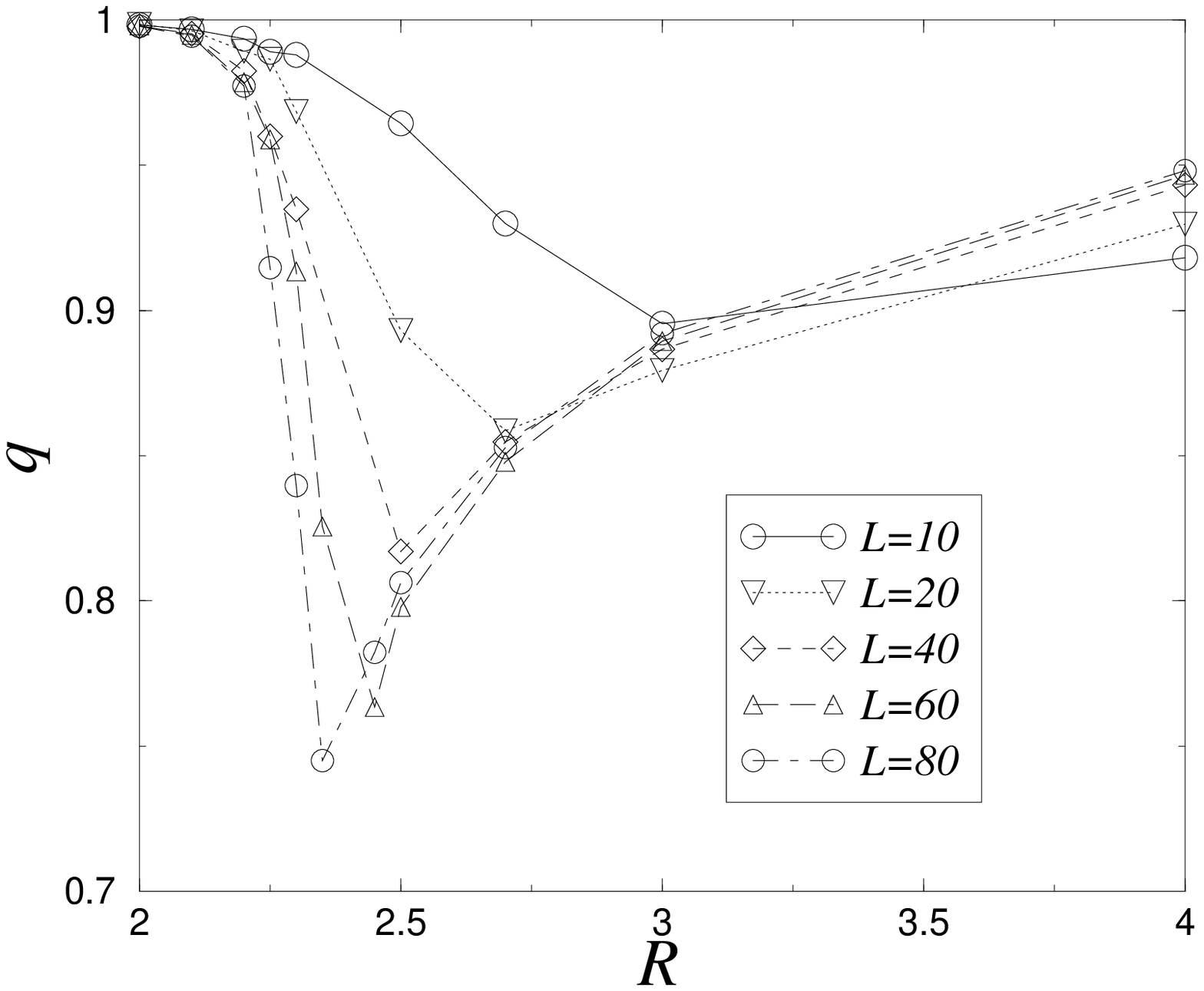,width=8cm,clip=!}}
\caption{The overlap between the GS and the DS in $d=3$ for different
system sizes.}
\label{figq}
\end{figure}

\begin{figure}[ht]
\centerline{\psfig{file=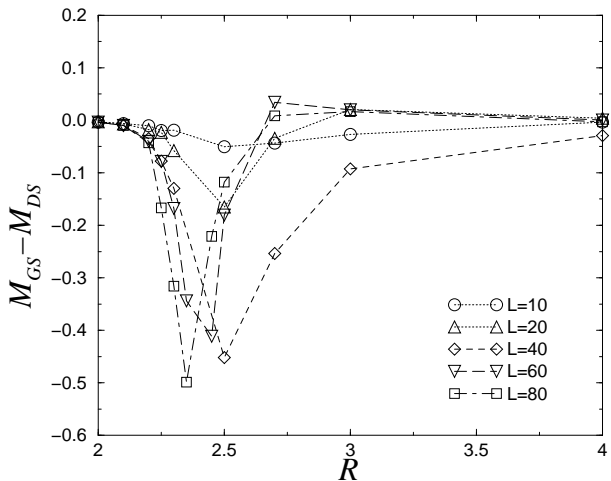,width=8cm,clip=!}}
\caption{The difference in magnetization between  
the GS and the DS in $d=3$ for different system sizes.}
\label{figdm}
\end{figure}


Here we analyze the problem again by numerical simulations, computing
the GS and the DS numerically, using the same disorder realizations
for the two cases. Simulations are performed for cubic lattices of
linear sizes $L=10,20,40,60,80$ and the results are averaged over
several realizations of the random fields. The GS is found exactly
using a min-cut/max-flow algorithm, while demagnetization is performed
approximately with the algorithm discussed in Ref.~\cite{DAN-02} with
$dH =10^{-3}$ (see section II). In both cases, we compute the average
magnetization as a function of the disorder width (see
Fig.~\ref{figmgsds}). In Fig.~\ref{figcollapse} we collapse the two
sets of data into a single curve, using two different values for $R_c$
(i.e. $R_c^{(GS)}=2.28$ and $R_c^{(DS)}=2.16$) but the same values for
the exponents (i.e. $1/\nu=0.73$ and $\beta=0.03$). The best value for
the ratio of the non-universal constant is found to be
$A_{DS}/A_{GS}\simeq 1$ and $B_{DS}/B_{GS} = 0.68 \pm 0.02$.  The fact
that the scaling function is the same for the two cases is a strong
indication for universality, going beyond the simple numerical
similarity of the exponents.  There is always the possibility that in
the limit $L\to \infty$ $R_c^{(GS)}=R_c^{(DS)}$. At the present stage
this hypothesis is not supported by the data, since we were not able
to collapse all the data into a single curve using the same $R_c$.

Next, we compare the statistical properties of the GS and the DS
around the transitions. In Fig.~\ref{figq} we report the value of the
overlap as a function of $R$ for different system sizes.  When the
disorder is decreased from the paramagnetic region, the overlap
decreases as for $d=1$. However for low disorder the overlap rapidly
increases and reaches $1$ in the ferromagnetic state. The minimum of
the overlap is located in the parameter region corresponding to the
transitions (i.e. $R\sim 2.2 - 2.3$).  A decrease in the overlap
around the transition can be expected, since for
$R_c^{(DS)}<R<R_c^{(GS)}$ the GS is ferromagnetic ($M>0$) and the DS
is paramagnetic ($M=0$) as it is also apparent plotting the difference
in the magnetization (see Fig.~\ref{figdm}). 

In summary, three dimensional simulations indicate that the
transitions in the GS and DS are universal, but the critical parameter
seems to differ. Consequently the GS and DS differ mostly around the
transition, while the difference is smaller in the paramagnetic and
ferromagnetic phases.

\subsection{The Bethe lattice}
The RFIM can be solved exactly in the Bethe lattice, displaying a
disorder induced transition in the GS and in the DS \cite{COL-02}.
It is thus an interesting case to compare the two states around the
respective transition directly in the thermodynamic limit.
We consider here a Bethe lattice with coordination $z$ and
obtain the GS  generalizing the $d=1$ case as in Ref. \cite{Bethe}.
In this case $N$ refers to the generation of the lattice, 
and $Z^{\pm}_n$ are the partition functions of a branch 
of generation $n$ with a fixed up (down) spin at the central site. 
The recursion relation for the $Z^{\pm}_n$ is  
\begin{equation}
Z_n^{\pm}(i)=e^{\pm\beta h_i}\prod_{j \in I(i)}
(Z_{n-1}^{+}(j)e^{\pm\beta J}+Z_{n-1}^{-}(j)e^{\mp\beta J})
\label{Z_nb}
\end{equation}
where for any given site $i$ the sum over $j$ runs over 
the $z-1$ nearest neighbors 
of $i$ away from the center of the lattice.
Then, following the $d=1$ case, one can write  
\begin{equation} 
F_n(i)\!=\!\!\!\!\sum_{j\in I(i)} \!\!\!
\left[F_{n\!-\!1\!}(j)\!-\!\frac{1}{2\beta}\ln 2\left(\cosh(\beta J)+\cosh(2 \beta x_n(j))\right)\right]
\label{Fb}
\end{equation}
where 
\begin{equation}
x_n(i)=\frac{1}{2 \beta} \ln (Z_n^+(i)/Z_n^-(i)),
\label{xb}
\end{equation}
so that the contribution at the free energy from site $i$ is 
\begin{equation} 
f(i)=-\frac{1}{2\beta}\ln(2\cosh\beta J+2 \cosh(2 \beta x_n(i))) .
\label{fb}
\end{equation}
$x_n(i)$ is a stochastic quantity satisfying the equation
\begin{equation} 
x_n(i)=h_i+\sum_{j\in I(i)} g(x_{n-1}(j))
\label{x2b}
\end{equation}
When $R \rightarrow 0$ Eq. (\ref{x2b}) has a 
fixed point solution of $x_\infty=(z-1)g(x_\infty)$. 
$x_\infty=0$ is a solution for any $J$ and $\beta$. 
For $\beta< \beta_c =\frac{1}{2} \ln \frac{z}{z-2}$ 
there are also two stable solutions $\pm x_{\infty} \ne 0$ 
corresponding to the appearance of a ferromagnetic phase.

\begin{figure}[ht]
\centerline{\psfig{file=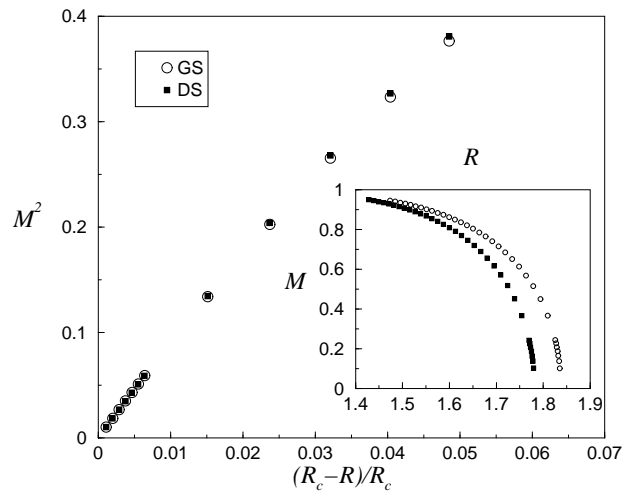,width=8cm,clip=!}}
\caption{The magnetization of the GS and the DS computed exactly on
the Bethe lattice with $z=4$ in the thermodynamic limit, showing the
ordering of the critical point (see inset).  When the data are plotted
against the reduced parameter $(R_c-R)/R_c$ the curves
superimpose. The result implies that for the Bethe lattice
$A_{GS}=A_{DS}$.}
\label{figbethe}
\end{figure}
 
To perform quenched averages one has to solve for the 
probability distribution 
of $W_n(x_n)$, where $W_n(x) dx=\mbox{Prob}(x<x_n<x+dx)$, which satisfies  
the recursive functional equation
 \begin{eqnarray}
&&W_{n+1}(x)=\int_{-\infty}^{\infty} dh P(h)
\int_{-\infty}^{\infty} dx_1 W_n(x_1)\cdots
\nonumber \\
&&
\cdots \!\int_{-\infty}^{\infty} \!\!\! dx_{z-1} W_n(x_{z\!-\!1})
\delta(x-\!h-\!H-\!\!\!\sum_{k=1}^{z-1}\!g(x_k)) \,,
\label{W_Nb}
\end{eqnarray}
so that in the thermodynamic limit $W_{\infty}$ is given by the fixed point equation 
 \begin{eqnarray}
&&
W_{\infty}(x)=
\int_{-\infty}^{\infty} dx_1 W_{\infty}(x_1) \cdots
\nonumber \\
&&
\cdots \!\int_{-\infty}^{\infty} \!\!\!  dx_{z\!-\!1}\! W_{\infty}(x_{z\!-\!1}) 
P(x\!-h\!-H-\!\!\!\sum_{k=1,}^{z-1}\!g(x_k)) .
\label{W_infb}
\end{eqnarray}
Once $W_{\infty}$ is known, any thermodynamic quantity can be computed. In particular, the free 
energy per spin is given again by (\ref{<f>})
and the magnetization at the central site of an infinite lattice, 
is given by Eq. (\ref{s0}) where $Z^{\uparrow\downarrow}$ 
are respectively the partition function with the spin 
at $0$ fixed up (down). 
These are given by 
\begin{equation} 
Z^{\uparrow\downarrow}=e^{\pm \beta h_0}
\prod_{k=1,z}
(e^{\pm \beta J}Z_k^+ +e^{\mp \beta J}Z_k^-)
\label{Zupdownb}
\end{equation}
and $Z_k^{\pm}$ for $k=1,\cdots z$ are the partition functions of the
$z$ branches attached to the central site $0$, with the boundary spin
fixed up(down). This gives for the magnetization at the central site 
$\langle s_0 \rangle$
\begin{equation} 
\langle s_0 \rangle =
\tanh(\beta(h_0+\sum_{k=1,z}g(x_k)))
\label{s02b}
\end{equation}
The magnetization for the infinite lattice can then be obtained averaging 
over the quenched variables $x_{r,l}$:
\begin{eqnarray}
&&
M=\int_{-\infty}^{\infty} dh P(h)
\int_{-\infty}^{\infty} dx_1 W_N(x_1) \cdots
\nonumber \\
&&
\cdots\int_{-\infty}^{\infty} dx_z W_N(x_z)
\tanh(\beta(h+\sum_{k=1,z}g(x_k))).
\label{magb}
\end{eqnarray}
For a Gaussian random-field distribution the fixed point equation
can not be solved explicitly and we thus resort to a numerical
integration. We obtain $W_{\infty}(x)$ for $z=4$, and for different values of 
$R$, and compute the magnetization using Eq.~(\ref{magb}).
In Fig.~\ref{figbethe} we compare the magnetization of the GS with
the one of the remnant magnetization in the DS, computed in 
Ref.~\cite{COL-02}. 
As observed in the simulations in $d=3$, the transition 
occurs at two different  
locations (see the inset of Fig.~\ref{figbethe}), for $z=4$
$R_c^{(DS)}=1.781258...$ \cite{COL-02} and $R_c^{(GS)}\simeq 1.8375$, with
the mean-field exponent ($\beta=1/2$). When plotted against 
$(R-R_c)/R_c$ the two curves superimpose close to the critical point. 
This indicates that, though not required by universality, in the Bethe 
lattice $A_{GS}=A_{DS}$, as also found in $d=3$.  

\begin{figure}[h]
\centerline{\psfig{file=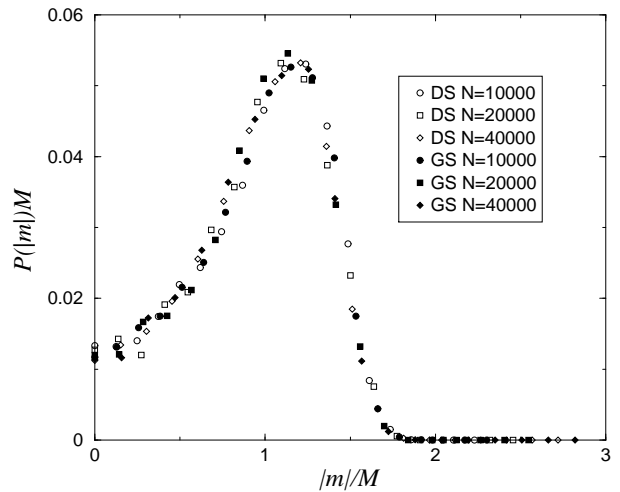,width=8cm}}
\caption{The distributions of the magnetization in the DS and the GS 
at their respective critical points on the Bethe lattice, obtained 
numerically for different lattice sizes $N$, can be all collapsed together.}
\label{figbethe2}
\end{figure}

To investigate possible
finite size scaling we have performed numerical simulations in the
Bethe lattice, following the method of Ref.~\cite{DHA-97}.  Collapsing
the order parameter curve as in $d=3$, using a scaling form similar to
Eq.~(\ref{eqfss}), does not appear to be possible in the Bethe
lattice, because the scaling region is very narrow.  Thus to test
finite size scaling, we have computed the distribution of the 
magnetization $m$ at the respective critical point, $R_c^{(DS)}$ and
$R_c^{(GS)}$ for different lattice sizes $N$.  The distributions can
all be collapsed into the same curve (see Fig.~\ref{figbethe2}), using
the form $P(|m|)=f(|m|/M)/M$.

\section{Reaching the ground state by non-equilibrium dynamics}

After having shown that the GS and the DS correspond to different
microscopic configurations, we investigate now if the GS spin
configuration may be reached by a field history other then the
ac-demagnetization. The answer to this question requires a
clarification on the relation existing between locally stable states
(given by the solutions of Eq.(\ref{EQ:2})), and the spin configurations
visited along the non-equilibrium dynamics induced by the varying
field. In fact, not all stable configurations may be reached by a
field history from saturation. The problem has been treated in
\cite{Sal} where it has been shown that, given a spin configuration
obtained by a field history, supposed unknown, the sequence of
reversal fields that applied to saturation gives back the original
state can be recovered. For spin systems this inverse function is given
by an algorithm which is able to construct the reverse field history
(RFH) \cite{Sal}. This method is applied then to investigate if a
given spin configuration may be reached by field history from
saturation: if a field history leading to the state exists the
algorithm produce a sequence of reversal fields; if no field history
exists the algorithm enters a recursive loop. The investigation of the
properties of unreachable states has been recently performed and leads
to a classification of stable states on oriented graphs
\cite{JEMS}. The study is performed here for the GS spin configuration
that, for the RFIM at finite size and for a given disorder
realization, can be independently derived by exact combinatorial
algorithms (as the \emph{max-flow min-cut}).

\subsection{RFH Algorithm}

Consider the final spin configuration $\mathbf{s}$ (the set of $N$
Ising spins) resulting after the application of a field history ending
at $H=0$ and consisting in a sequence of reversal fields $\{H\}=\{H_1,
\dots ,H_n\}$ from the saturated state and let us define the function
$\mathbf{s}=f(\{H\})$. The set of all states obtained this way is
defined as the hysteresis states (H-states). Due to adiabatic
dynamical response and return point memory, this state $\mathbf{s}$
will contain the memory of a subset of the reversal fields. In fact
not all the reversal fields determine the final state
$\mathbf{s}$. For example, in terms of average magnetization, the
reversal fields which give rise to closed minor loops do not influence
the final state, i.e. their memory is erased, while the memory of the
set of reversal fields $\{H_S\}$ which are not erased is contained in
the final state. The inverse function $\{H_S\}=g(\mathbf{s})$ allows,
starting from a spin configuration $\mathbf{s}$ belonging to the
H-states ensemble, to obtain the set of reversal fields $\{H_S\}$
which have been actually stored in the state and that - if applied as
a field history - will reproduce the original state,
i.e. $\mathbf{s}=f( g(\mathbf{s}) )$. We define this set of reversal
fields $\{H_S\}$ as \emph{minimal field history}.

The RFH algorithm takes as input a configuration $\mathbf{s}$ at $H=0$
and gives as output - when it exists - the reversal field history from
saturation to the state $\mathbf{s}$. The formulation of the algorithm
is based on the order-preserving character of the dynamics
\cite{SET-93}, and is therefore, applicable to a wide range of models
beyond the RFIM. An interesting result of the RFH algorithm is
obtained when it is applied to a state $\mathbf{s}$ not belonging to
the H-states (i.e. where no field history exists). The iterated search
for the reversal field sequence enters an iteration and, in this case,
it can be shown that no field history leading to the state exists.

\subsection{Simulation results in 1d}

\begin{figure}[htb]
\narrowtext 
\centering
\includegraphics[width=7.5cm,angle=-90]{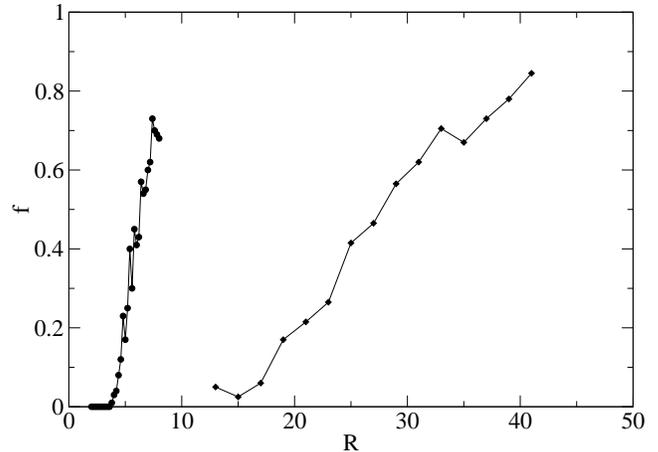}
\caption{ 
Fraction of reachable states (averages over 100 realizations of disorder) diamonds: 
fraction $f_{RND}$ ; circles: fraction $f_{GS}$
} 
\label{cap:fH}
\end{figure}

The RFH algorithm was applied to explore the possibility to reach the
GS by non equilibrium dynamics by the numerical study of the RFIM in
one dimension with periodic boundary conditions. We performed our
investigations on systems having $N=5000$ spins, averaging the results
for 100 different realizations of the same disorder $R$. The GS was
obtained by the \emph{max-flow min-cut} procedure for each realization
and the RFH algorithm was applied. At each disorder value $R$, the
fraction $f_{GS}$ of the realizations in which the GS resulted to be
reachable was computed. For comparison the same procedure was applied
starting from locally stable states generated by random sampling the
set of local minima. The results are shown in Fig.\ref{cap:fH}.

As a first finding the GS does not result to be systematically field
reachable and the fraction depends on the disorder. One may conclude
that the fact that the GS is sometimes reachable is a pure effect of
the finite system size. However, also for the random states the
fraction of found states f$_{RND}$ sensibly changes with $R$, but
following a different curve. If there was no correlation between GS
and the H-states the two curves would be coincident. The dependence of
f$_{RND}$ on $R$ reflects the fact that the number of H-states depends
on the disorder value and the system size \cite{Hstates}, and only at
large disorder, where the number of locally stable states decreases,
the ratio between H-states and stable states is significantly greater
then zero.

\section{conclusions}

For disordered systems like the random field Ising model
one would be interested in both universality in statistical
properties and in the question how to ``optimize'' in the
case of a sample with a given distribution of the impurities.
In this paper we have studied this problem in detail, by
comparing the demagnetized and ground states. Our main
findings are the following: First, the character of the GS
is such that it is globally optimized, and the demagnetization
procedure does not perform well unless the optimization problem
is rather trivial. This is slightly surprising since the conclusion
holds in particular if the RFIM GS is paramagnetic. Then the
DS does not manage to find the right spin configuration, so
that as seen in the $d=1$ case many of the domains of the GS
do not appear in the DS. 

Second, in $d=3$ (and with the aid of the Bethe lattice solution),
it can be demonstrated that the existence of a phase transition
for both the DS and GS makes the ``phase diagram'' of optimization
to show a regime where the outcome is less optimal: in
the paramagnetic phase of the DS, where the GS is already ferromagnetic
since the critical thresholds are ordered such that
$R_c^{(GS)}>R_c^{(DS)}=1.84$. In this regime DS and GS are expected to
differ strongly in the thermodynamic limit. We also
provide numerical evidence that the $d=3$ transition
appears to have the same critical exponents in both the GS and DS
\cite{nota-perez}.
This can be considered both surprising -- there being no exact
field theoretical way of treating the $d=3$ phase transition --
and expected, since the functional renormalization calculations
in spite of their shortcomings indicate that the actions are the
same \cite{DAH-96}. It seems intriguing that such universality is met exactly
in the limit where the ``optimized'' character of the DS changes. 

The results indicate that for the particular system at hand,
where the disorder couples directly to the expected
magnetization, ``local'' optimization methods have difficulties.
Of course, as in ``hysteretic optimization'', one can perturb
or ``shake'' the state obtained from the DS procedure to try to 
still lower the energy. These attempts are of course usually
just heuristic. In the case of the RFIM, the joint approach
of optimizing by the DS and computing the GS exactly allows
to understand better similarities and differences between
equilibrium and low energy non-equilibrium states.

In addition to the ferromagnetic RFIM model, one
can consider other systems where two disorder induced phase 
transitions exist. Numerical simulations and analytical results have
shown that a disorder induced transition in the hysteresis loop can be
observed in the random bond Ising model \cite{VIV-94}, in the random
field O(N) model \cite{DAS-99}, in the random anisotropy model
\cite{VIV-01,DAS-04} and in the random Blume-Emery-Griffith model
\cite{VIV-94}. All these systems display as well a transition in equilibrium
and it would be interesting to compare their DS and GS. 

Interfaces in quenched disorder would provide another interesting
example, since the roughness exponent typically differs in and out of
equilibrium (i.e. at the depinning threshold)
 \cite{NAT-00}.  It would be interesting to measure the roughness of
an interface after a demagnetization cycle (i.e. after the field 
driving the interface is cycled with decreasing amplitude), and compare 
its properties with those of the ground state interface. 
Finally,  there is the issue of energetics of excitations
in the respective ensembles: the universality of exponents
and scaling functions would seem to imply that these also
scale similarly.


\begin{thebibliography}{10}
\bibitem{ALA-01}
M. Alava, P. Duxbury, C. Moukarzel, and H. Rieger,  
in {\em Phase transitions
  and critical phenomena, Vol 18}, edited by C. Domb and J. Lebowitz (Academic
  Press, San Diego, 2001).

\bibitem{Bertotti}
G. Bertotti, {\em Hysteresis in Magnetism} (Academic Press, San Diego, 1998).

\bibitem{ZAR-02}
G. Zarand, F. Pazmandi, K. F. Pal, and G. T. Zimanyi,
Phys. Rev.  Lett. {\bf 89}, 150201 (2002).

\bibitem{NAT-00}
T. Nattermann, in  
{\em  Spin Glasses and Random Fields} edited by A.P. Young 
(World Scientific, Singapore,  1997).


\bibitem{OGI-86}
A. T. Ogielski, Phys. Rev. Lett. {\bf 57}, 1251 (1986).

\bibitem{HAR-99}
A. K. Hartmann and U. Nowak, 
Eur. Phys. J. B {\bf 7}, 105 (1999).

\bibitem{HAR-01}
A. K. Hartmann and A. P. Young 
Phys. Rev. B {\bf 64}, 214419 (2001).
 
\bibitem{MID-02}
A.~A. Middleton and D.~S. Fisher,
Phys. Rev. B {\bf 65}, 134411 (2002).


\bibitem{BRU-84}
R. Bruinsma,  Phys. Rev. B {\bf 30}, 289  (1984).
 
\bibitem{SWI-94}
M.~R. Swift, A. Maritan, M. Cieplak, and J.~R. Banavar, 
J. Phys. A {\bf 27},  1525  (1994).

\bibitem{SLA-99}
Z. Slanic, D. P. Belanger, and J. A. Fernandez-Baca
Phys. Rev. Lett. {\bf 82}, 426 (1999).
                                                                                              
\bibitem{YE-02}
F. Ye et al., Phys. Rev. Lett. {\bf 89}, 157202 (2002).

\bibitem{SET-93}
J.~P. Sethna et. al, Phys. Rev. Lett. {\bf 70},  3347  (1993)

\bibitem{DAH-96}
K. Dahmen and J.~P. Sethna, Phys. Rev. B {\bf 53},  14872  (1996).

\bibitem{PER-99}
O. Perkovic, K.~A. Dahmen, and J.~P. Sethna, Phys. Rev. B {\bf 59},  6106
  (1999).

\bibitem{SET-01}
J.P. Sethna, K.~A. Dahmen, and C.~R. Myers,
Nature {\bf 410} 242 (2001).

\bibitem{BER-00}
A. Berger, A. Inomata, J.~S. Jiang, J.~E. Pearson, and S.~D. Bader, Phys. Rev. 
Lett. {\bf 85}, 4176 (2000). 

\bibitem{MAR-03}
J. Marcos, et. al. Phys. Rev. B {\bf 67}, 224406 (2003).

\bibitem{PER-03}
F. J. P\'erez-Reche and E. Vives
Phys. Rev. B {\bf 67}, 134421 (2003).

\bibitem{PER-04}
F. J. P\'erez-Reche and E. Vives
 preprint cond-mat/0403754.

\bibitem{DAN-02}
L. Dante, G. Durin, A. Magni, and S. Zapperi, Phys. Rev. 
B {\bf 65}, 144441 (2002). 

\bibitem{COL-02}
F. Colaiori, A. Gabrielli, and S. Zapperi,  Phys. Rev. B {\bf 65}, 
224404 (2002). 

\bibitem{SHU-96}
P. Shukla, Physica A {\bf 233},  235  (1996).

\bibitem{SHU-00}
P. Shukla, Phys. Rev. E {\bf 62},  4725  (2000).

\bibitem{DHA-97}
D. Dhar, P. Shukla, and J.~P. Sethna, J. Phys. A {\bf 30},  5259  (1997).

\bibitem{SAB-00}
S. Sabhapandit, P. Shukla, and D. Dhar, J. Stat. Phys. {\bf 98},  103  (2000).

\bibitem{SHU-01}
P. Shukla, Phys. Rev. E {\bf 63},  027102  (2001).

\bibitem{MAR-94}
A. Maritan, M. Cieplak, M.~R. Swift, and J.~R. Banavar, Phys. Rev.
Lett. {\bf 72},  946  (1994)
                                                                                              
\bibitem{SET-94}
J.~P. Sethna {\em et al.}, Phys. Rev. Lett. {\bf 72},  947  (1994)

\bibitem{SOU-97}
J.-C. Angl\'s d'Auriac and N. Sourlas
Europhys. Lett. {\bf 39}, 473 (1997). 

\bibitem{DUX-01}
P.~M. Duxbury and J.~M. Meinke
Phys. Rev. E {\bf 64}, 036112 (2001).

\bibitem{DOB-02}
R. Dobrin, J.~M. Meinke, and P.~M. Duxbury,
 J. of Phys. A: Math. Gen. {\bf 35}, L247 (2002).

\bibitem{CAR-01}
J.~H. Carpenter et al. , J. Appl. Phys. {\bf 89},  6799  (2001).

\bibitem{COL-04}
F. Colaiori, M. J. Alava, G. Durin, A. Magni, and S. Zapperi, 
 Phys. Rev. Lett. {\bf 92}, 257203 (2004).

\bibitem{1drfim}
G. Schr\"oder, T. Knetter, M. J. Alava and H. Rieger, Eur. Phys. J. B {\bf 24},
101-105 (2001).

\bibitem{eira}
E. Sepp\"al\"a, PhD thesis, HUT, Helsinki, 2001.

\bibitem{BER-90}
G. Bertotti and M. Pasquale, J. Appl. Phys. {\bf 67}, 5255 (1990).

\bibitem{Bethe}
R. Bruinsma, Phys. Rev. B, {\bf 30}, 289 (1984). 

\bibitem{CAR-03}
J. H. Carpenter and K. A. Dahmen
Phys. Rev. B {\bf 67}, 020412 (2003) 

\bibitem{Sal}
V. Basso and A. Magni, Physica B {\bf 343}, 275 (2004).

\bibitem{JEMS}
A. Magni and V. Basso, submitted to JEMS 04 Conference, (J. Magn.Magn.Mat.).

\bibitem{Hstates}
V. Basso and A. Magni, to be published.

\bibitem{nota-perez}
A similar suggestion was made independently in Ref.~\cite{PER-04}
\bibitem{VIV-94}
E. Vives and A. Planes, Phys. Rev. B {\bf 50},  3839  (1994).
 

\bibitem{DAS-99}
R. da~Silveira and M. Kardar, Phys. Rev. E {\bf 59},  1355  (1999).
                                                                                              
\bibitem{VIV-01}
E. Vives and A. Planes, Phys. Rev. B {\bf 63},  134431  (2001).

\bibitem{DAS-04}
R. da~Silveira and S. Zapperi, Phys. Rev. B {\bf 69}, 212404 (2004) 

\end{thebibliography}
\end{document}